# Evidence for two electronic components in high-temperature superconductivity from NMR

J Haase[1], C P Slichter[2] and G V.M. Williams[3]

[1]University of Leipzig, Faculty of Physics and Earth Science, 04103 Leipzig, Germany

[2]University of Illinois at Urbana-Champaign, Urbana, IL 61801, USA

[3]The MacDiarmid Institute, Industrial Research, P.O. Box 31310, Lower Hutt, New Zealand

Email: j.haase@physik.uni-leipzig.de

**Abstract**. A new analysis of $^{63}Cu$ and $^{17}O$ NMR shift data on $La_{1.85}Sr_{0.15}CuO_4$ is reported that supports earlier work arguing for a two-component description of $La_{1.85}Sr_{0.15}CuO_4$, but conflicts with the widely held view that the cuprates are a one-component system. The data are analyzed in terms of two components A and B with susceptibilities $\chi_{AA}$, $\chi_{AB} (= \chi_{BA})$, and $\chi_{BB}$. We find that above $T_c$, $\chi_{AB}$ and $\chi_{BB}$ are independent of temperature and obtain for the first time the temperature dependence of all three susceptibilities above $T_c$ as well as the complete temperature dependence of $\chi_{AA} + \chi_{AB}$ and of $\chi_{AB} + \chi_{BB}$ below $T_c$. The form of the results agrees with that recently proposed by Barzykin and Pines.

PACS numbers: 74.72.-h, 76.60.-k, 74.25.Jb

Soon after the discovery of high-temperature superconductivity (Bednorz and Müller, 1986), the issue arose whether the system needed one or two components to describe the low energy magnetic properties. There was agreement that the parent antiferromagnet is a Mott insulator (Anderson, 1987), and that the CuO$_2$-plane consists of magnetic Cu in $3d^9$ configuration with a hole in the $d(x^2-y^2)$ orbital hybridized with O $2p_\sigma$ orbitals of the four surrounding, nearly closed shell oxygen $2p^6$ ions. However, experiments (Fujimori et al., 1987; Nücker et al., 1987; Tranquada et al., 1987) showed that hole doping mainly affects the $2p_\sigma$ orbitals (Haase et al., 2004). While this may favor two-component approaches (Castellani et al., 1988; Emery, 1987; Gor'kov and Sokol, 1987), it was suggested early on by Zhang and Rice (Zhang and Rice, 1988) that a single-band effective Hamiltonian can be appropriate if the oxygen holes form stable singlets with the central Cu. Mila and Rice (Mila and Rice, 1989a) showed that the NMR data of the planar Cu in $YBa_2Cu_3O_{7-y}$ could be explained with Cu moments only, and later argued (Mila and Rice, 1989b) that Y NMR data (Alloul et al., 1989) support a single-fluid model. While there were early attempts in interpreting the NMR data in terms of two-component scenarios, e.g. (Cox and Trees, 1990), when Takigawa et al. (Takigawa et al., 1991) reported that planar Cu and O shifts in $YBa_2Cu_3O_{6.63}$ were approximately proportional to the uniform spin susceptibility, their account was taken by many as proof for the validity of a single-fluid picture for high-temperature superconductivity. This assumption supported the quite successful Millis-Monien-Pines model (Millis et al., 1990) of the spin susceptibility that explained many NMR properties very well (but did have difficulties (Zha et al., 1996) with accounting for the incommensurate peaks observed with neutron scattering). Later, Walstedt (Walstedt et al., 1994) argued on the basis of relaxation measurements of planar Cu and O in $La_{1.85}Sr_{0.15}CuO_4$ that a single-fluid scenario was not appropriate for this material, as suggested by Johnston (Johnston, 1989) who showed that the uniform spin susceptibility could be decomposed into two terms. His analysis was confirmed by Nakano et al. (Nakano et al., 1994) later on.

Recently, we have performed a more rigorous analysis of the spin shifts for $La_{1.85}Sr_{0.15}CuO_4$ (Haase et al., 2008) and found that the results were in disagreement with the response of a single electronic fluid. Here we present more details and a new analysis that, firstly, underscores the significance of the failure of the single-component description (as we can relax the assumption of a vanishing spin shift at low temperatures, which is usually adopted). Second and more importantly however, our new analysis shows that our first analysis (Haase et al., 2008) is in general not appropriate as we neglected a third term for the susceptibilities of a two-component system, which was e.g. introduced by Curro, Young, Schmalian and Pines (Curro et al., 2004) for the description of heavy-electron materials. We now find that the previously neglected term $\chi_{AB}$ (see below) that is due to the coupling *between* the two components *A* and *B* is indeed present and plays an important role.

We now begin with the new analysis of our experimental data. We will find that the form of the resulting analysis is similar to that recently proposed by Barzykin and Pines (Barzykin and Pines, 2009).

For a single electronic fluid the anisotropic NMR spin shift can be written as

$$K_k(T) = p_k \chi(T), \quad p_k = \frac{h_k}{\gamma_k \gamma_e \hbar^2}, \tag{1}$$

where $h_k$ is the orientation-dependent magnetic hyperfine constant, $\gamma_k$ and $\gamma_e$ are the gyromagnetic constants for the nucleus $k$ and the electron, respectively, and $\chi(T)$ is the temperature-dependent uniform spin susceptibility (which we consider to be isotropic). If $La_{1.85}Sr_{0.15}CuO_4$ was a single-fluid material, then we would expect the spin shifts at all nuclei for all orientations of the external field with respect to the crystal axes to follow (1). (Note that $YBa_2Cu_3O_{7-y}$ could have two components from planes and chains.) This means that at different nuclear sites the changes in spin shift between any two temperatures $T$ and $T_0$, $\Delta K_k = K_k(T) - K_k(T_0)$ would have to be proportional to the single-fluid's change in spin susceptibility $\Delta\chi = \chi(T) - \chi(T_0)$ between these any two temperatures (we can let $k$ denote both the nuclear site and the orientation of the crystal c-axis with respect to the external field for which the shift has been measured; note that for a particular $k$ the shift difference could be zero if the corresponding hyperfine constant vanishes). Such NMR spin shift measurements in high magnetic fields $B_{external}$ are difficult because there is a temperature-dependent, anisotropic Meissner shift (Pennington et al., 1989) $K_{M,k}(T)$ below the superconducting transition temperature $T_c$. Thus, the experimentally measured shift is given by,

$$K_{exp,k}(T) = K_k(T) + K_{M,k}(T). \qquad (2)$$

Note that quadrupolar shifts and shifts from core and bonding electrons are temperature independent and therefore do not interfere with the analysis.

For the further analysis of our experimental data we label the shifts as follows: We use numbers to label nuclei and magnetic field orientation. *1* and *2* denote $^{63}Cu$, *3* and *4* denote planar $^{17}O$, *5* and *6* denote the apical oxygen. For *1,3,5* the magnetic field $B_{external}$ is parallel to the c-axis, for *2,4,6* it is perpendicular to the c-axis. The planar Cu shift, $K_1$, for $B_{external} \parallel c$ is independent of $T$ and doping, and the planar O shift, $K_3$, for $B_{external} \perp c$ was not determined since the line is too broad for this orientation with the c-axis aligned sample.

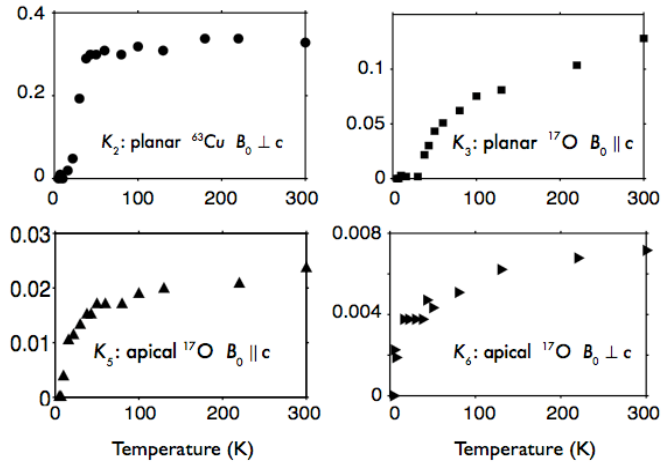

Figure 1: NMR spin shifts $K_n$ (%) as a function of temperature for various nuclei with the magnetic field $B_0 = 9$ Tesla perpendicular ($B_0 \perp c$) and parallel ($B_0 \parallel c$) to the crystal c-axis.

In figure 1 we show the *T*-dependent shift data defined as the difference between its value at temperature *T* and its value at $T \approx 0$. Note that the shifts include a possible Meissner term $K_{M,k}(T)$ that will not depend on the nuclear species, but may depend on the orientation of the external field with respect to the crystal c-axis since the vortex structure is anisotropic.

In order to probe single-component behavior we form the following experimental shift differences, cf. (2),

$$\Delta G_\perp \equiv \left[K_{\exp,2}(T) - K_{\exp,2}(T_0)\right] - \left[K_{\exp,6}(T) - K_{\exp,6}(T_0)\right] = \left[K_2(T) - K_2(T_0)\right] - \left[K_6(T) - K_6(T_0)\right],$$
$$\Delta G_\parallel \equiv \left[K_{\exp,3}(T) - K_{\exp,3}(T_0)\right] - \left[K_{\exp,5}(T) - K_{\exp,5}(T_0)\right] = \left[K_3(T) - K_3(T_0)\right] - \left[K_5(T) - K_5(T_0)\right].$$
(3)

Note that the Meissner terms disappear. Now, these shift differences must, for a single fluid, be proportional, cf. (1), to the difference of the susceptibility at the two temperatures, so that $\Delta G_\perp = c_\perp \left[\chi(T) - \chi(T_0)\right]$, $\Delta G_\parallel = c_\parallel \left[\chi(T) - \chi(T_0)\right]$, where $c_{\perp,\parallel}$ are constants. Consequently, for a single-component system we must have

$$\Delta G_\perp = \frac{c_\perp}{c_\parallel} \Delta G_\parallel.$$
(4)

The corresponding experimental plot is shown in figure 2 (left). It is obvious that the linear response of the system (independent of any assumption about zero shift and zero susceptibility) cannot be described by a single component's susceptibility. From the plot we find an approximate linear relationship $\Delta G_\perp(T > T_c) = \frac{c_\perp}{c_\parallel} \Delta G_\parallel(T > T_c) + const.$, and we estimate $c_\perp / c_\parallel \approx 0.38$ and $\Delta G_\perp(\Delta G_\parallel = 0) \approx 2.85 \cdot 10^{-3}$. While the temperature above which both terms are proportional to each other seems to coincide with the superconducting critical temperature $T_c$, we do not know whether this is indeed the case or just accidental. We therefore prefer to call this temperature $T_{const.}$ and we find with our data that we cannot distinguish it with certainty form $T_c$.

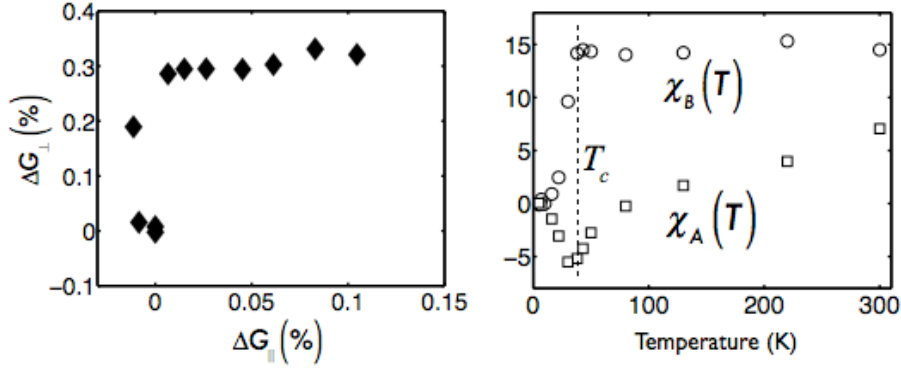

Figure 2: Left: Plot of the measured shift difference $\Delta G_\perp$ (planar Cu minus apical O; magnetic field perpendicular to the crystal c-axis: $B_0 \perp c$) as a function of $\Delta G_\parallel$ (planar O minus apical O; with $B_0 \parallel c$), cf. (3). The lack of proportionality is evident. Right: Temperature dependence of the two susceptibilities $\chi_A = \chi_{AA} + \chi_{AB}$ and $\chi_B = \chi_{BB} + \chi_{AB}$ (in units of $10^{-5} emu/mol$) that follow from the NMR spin shift data.

Since a single component description fails to explain our data we assume that each nuclear spin couples to *two* different electronic spin components with the two susceptibilities $\chi_A(T)$ and $\chi_B(T)$, so that we write instead of (1),

$$K_k(T) = p_k \chi_A(T) + q_k \chi_B(T), \qquad (5)$$

where $p_k$ and $q_k$ are the two generalized hyperfine coupling coefficients for a particular nucleus at given orientation of the sample with respect to the external magnetic field (denoted by the index $k$) to the two electronic spin components $A$ or $B$. At this point, to keep the analysis as general as possible, we do not specify the meaning of "$A$" or "$B$". Later, we see that "$A$" refers to the Cu electron spin and "$B$" refers to the planar oxygen electron spin.

In our previous paper (Haase et al., 2008), written in 2006, we assumed that these two susceptibilities must be the ones that had been found with magnetization measurements (Johnston, 1989; Nakano *et al.*, 1994) above $T_c$. This assumption was wrong, as we explain now. If we place a two-component system with the two fluids *A* and *B* in an external magnetic field the induced total magnetic moment $M_{total}$ will be given by $M_{total} = (\chi_{AA} + 2\chi_{AB} + \chi_{BB})B_{external}$ so the uniform susceptibility $\chi_0$ is the sum of three terms(Curro et al., 2004), $\chi_0 = \chi_{AA} + 2\chi_{AB} + \chi_{BB}$.

The two terms $\chi_{AA}$ and $\chi_{BB}$ are the susceptibilities of the hypothetically isolated components *A* and *B*, respectively. The term $\chi_{AB} = \chi_{BA}$ is caused by the coupling between the two components *A* and *B*, and describes the electron spin polarization of the component *A* due to a spin polarization of component *B*, and vice versa. As a consequence, for example, a nuclear spin that has a hyperfine coupling directly to the electron spin of component *A* will measure the response $\chi_{AA}$ of component *A* due to the external field acting on *A*, as well as the response $\chi_{BA} = \chi_{AB}$ of *A* due to the external field acting on component *B*. With Equation (5) we then have,

$$K_k(T) = p_k \chi_A + q_k \chi_B \text{ with } \chi_A = \chi_{AA} + \chi_{AB}, \quad \chi_B = \chi_{BB} + \chi_{AB}.$$

We now proceed with the shift analysis. For any set of two shifts $K_k, K_l$ we can eliminate one susceptibility that we call $\chi_A$,

$$K_k(T) = \frac{p_k}{p_l} K_l(T) + \left\{ q_k - \frac{p_k}{p_l} q_l \right\} \chi_B(T). \tag{6}$$

In such an approach we have with (3) $c_\perp / c_\parallel = (p_2 - p_6)/(p_3 - p_5)$, and for the *T*-independent term $\Delta G_\perp (\Delta G_\parallel = 0) = \{ q_2 - q_6 - c_\perp / c_\parallel \cdot (q_3 - q_5) \} \chi_B(T > T_{const.}) \approx 2.85 \cdot 10^{-3}$.

We now plot all six pairs of T-dependent shifts (that include the Meissner term) one against the other. The result is shown in figure 3. We find that all plots are approximately linear at higher temperatures, as one expects from figure 2. We conclude that $\chi_B$ is independent of temperature above $T_{const.}$.

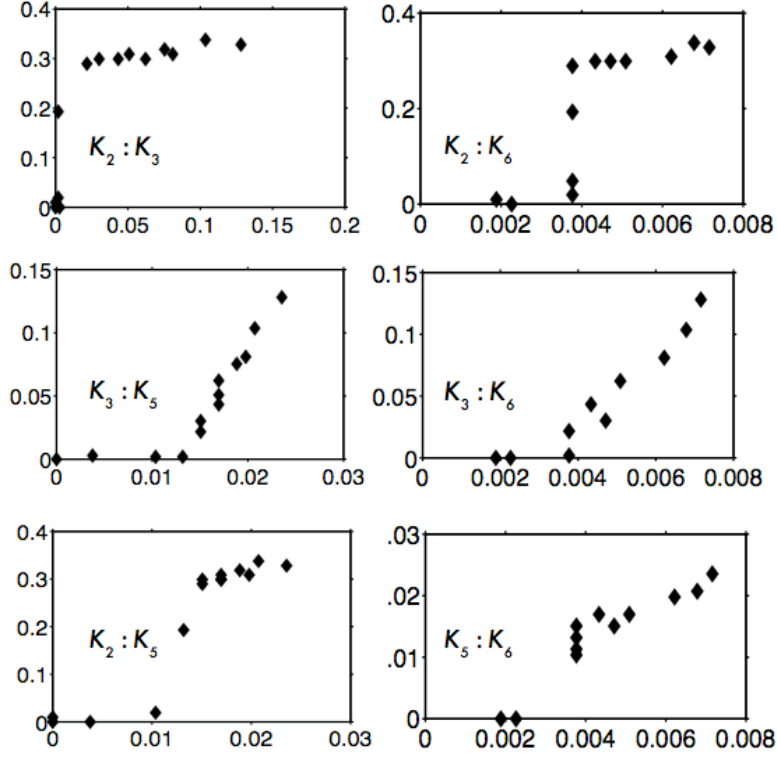

Figure 3: NMR spin shifts from figure 2 plotted against each other, linearity at higher temperatures is observed for all plots.

Since an upper limit to the Meissner shifts are given by the apical O shifts and since they are rather small, we neglect the Meissner terms momentarily. Then, we infer from (6) that at higher temperatures the sum $\chi_{BB} + \chi_{AB}$ is T-independent. It seems highly unlikely that both, $\chi_{AB}$ and $\chi_{BB}$ are T-dependent, and their sum is not. So we conclude that $\chi_{AB}$ and $\chi_{BB}$ are both temperature independent above $T_c$, and that $\chi_A(T > T_{const.}) = \chi_{AA}(T) + \chi_{AB}$ is the sum of the T-dependent $\chi_{AA}(T)$ and the T-independent $\chi_{AB}$. From the six plots in Figure 3 and with (6) we can determine all ratios $p_k / p_l \equiv s_{kl}$ $(s_{23} = 0.45, s_{24} = 5.0, s_{34} = 12.2, s_{36} = 33.3, s_{46} = 2.5)$, as well as all constants $q_k \chi_B \equiv \kappa_k$ $(\kappa_2 = 3.1, \kappa_3 = 0.67, \kappa_5 = 0.18, \kappa_6 = 0.055$, all in units of $10^{-3})$. With these numbers $\kappa_k$ we calculate $\{q_2 - q_6 - c_\perp / c_\parallel \cdot (q_3 - q_5)\} \chi_B (T > T_{const.}) \approx 2.86 \cdot 10^{-3}$, experimentally the same value as found earlier from the plot of $\Delta G_\perp$ vs. $\Delta G_\parallel$. This shows that it is legitimate to discard the Meissner terms for our analysis at higher temperatures.

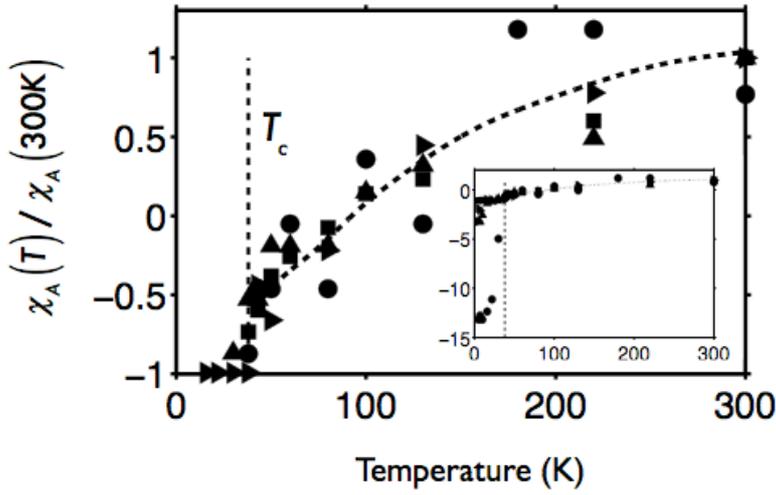

Figure 4: Inset: $\chi_A(T) / \chi_A(300K)$ as a function of temperature, as obtained from the four shift plots. The same symbol assignment for the shifts as in figure 2 was used here. The dashed line is Johnston's [19] $\chi_1(T)$, scaled and shifted vertically to fit the date. Main panel: Blow-up of the higher temperature part.

For the four experimental shifts, $K_2, K_3, K_5,$ and $K_6$, we can rearrange Eq. (5) to get four plots of

$$\chi_A(T) = [K_k(T) - \kappa_k] / p_k$$

normalized to its value at 300K. Note that such a plot should produce a unique function for all shifts $k$ above $T_{const.}$. This is indeed the case as figure 4 shows. (For Cu $K_2(300K)$ was determined from a fit to $K_2(T > T_{const.})$ to a straight line since the scatter is very large for Cu, as a large number $\kappa_2$ has been subtracted). One also observes in Figure 4 that the susceptibility $\chi_A$ changes sign near 100K, $\chi_A(T \approx T_{const.})/\chi_A(300K) < 0$. This means that $\chi_{AA}$ and/or $\chi_{AB}$ must be negative already above $T_{const.}$. Since $\chi_B$ is approximately constant above $T_{const.}$, we know that $\chi_{AB}(T > T_{const.}) \approx const.$, as well. On general ground one may argue that $\chi_{AA}$ should be positive at all temperatures so that a constant, but negative $\chi_{AB}(T > T_{const.})$ is the most likely explanation for the observed negative behavior of $\chi_A$.

Since our results demand that above $T_{const.}$ the uniform spin susceptibility must be given by $\chi(T > T_{const.}) = \chi_{AA}(T) + [2\chi_{AB} + \chi_{BB}]$, where only the first term is $T$-dependent, this must agree with the results of Johnston (Johnston, 1989) who found with magnetization measurements above $T_c$ that the spin susceptibility can be written as a sum of two terms, a T-*in*dependent term and a constant, but doping dependent term. These findings were verified by Nakano et al. (Nakano et al., 1994) later on. Using Johnston's notation, the spin susceptibility can be written as $\chi(x, T > T_{const.}) = \chi_2(x) + \chi_1(T)$, $\chi_1(T) = [\chi_m(x) - \chi_2(x)] F(T/T_{max})$, where $T_{max}(x)$ is the temperature at which the susceptibility has its maximum $\chi_m(x)$, and $F(T/T_{max})$ is a universal, doping independent function. The *T-in*dependent part he wrote as $\chi_2(x) = \chi_{core} + \chi_{VV} + \chi_p(x)$, where, in addition to a contribution from the core diamagnetism $\chi_{core}$ and a Van-Vleck term $\chi_{VV}$, a doping dependent term $\chi_p(x)$ is present that Johnston suggested stems from the doped holes' Pauli susceptibility (for $x = 0$ he consequently demanded $\chi_p(0) \equiv 0$). Comparing with our own results this means that $\chi_1(T) = \chi_A(T) + C_1$. We can test whether the temperature dependence of the spin shift is consistent with the function $F(T/T_{max})$ found by Johnston. To do this, we plot the Johnston function, scaled to fit our data between 300 K and $T_{const.}$. The resulting plot $\chi_{fit}(T)$, the dashed line in Figure 4, obeys the equation $\chi_{fit}(T) = 2.6\, F(T/T_{max}) - 1.5$.

So we have two *T*-dependent functions, $\chi_1$ of Johnston and our $\chi_{AA}$. In addition, we can express the total spin susceptibility two ways: $\chi_{spin} = \chi_1 + \chi_2$ or $\chi_{spin} = \chi_{AA} + 2\chi_{AB} + \chi_{BB}$. Both $\chi_2$ and $2\chi_{AB} + \chi_{BB}$ are independent of *T* above $T_{const.}$. To be consistent, $\chi_1$ can differ from $\chi_{AA}$ at most by an additive constant. Since $\chi_1$ obeys a universal scaling law, it seems most reasonable to us to assume that $\chi_{AA} = \chi_1$. We proceed on that assumption.

From Johnston's data we estimate for $x \approx 0.15$ $(T_{max} \approx 420K)$: $\chi_2 \approx 2.8 \cdot 10^{-5} \, emu/mol$ and $\chi_m \approx 10 \cdot 10^{-5} \, emu/mol$. With the doping independent contributions from core diamagnetism and the Van-Vleck term (Johnston, 1989) we calculate $\chi_p(x \approx 0.15) \approx 10.3 \cdot 10^{-5} \, emu/mol$. Consequently, we can estimate the three components to the susceptibility for $x \approx 0.15$ and find in units of $10^{-5} \, emu/mol$, $\chi_{AA}(T > T_c) \equiv \chi_1(T) \approx +7.2 \cdot F(T/T_m)$, $\chi_{AB} \approx -4.2$, $\chi_{BB} \approx +18.7$.

At 300K, $F(300K/T_{max}) \approx 0.98$ (Johnston, 1989), and we can thus determine $\chi_A(300K)$, $\chi_B(300K)$, and eventually the hyperfine coefficients of the nuclei with the two electronic spin components. We derive the following numbers (for two different units common in the literature), cf. (5):

$p_2 = 2.6$, $p_3 = 8.7$, $p_5 = 0.79$, $p_6 = 0.24$, $q_2 = 21.4$, $q_3 = 4.6$, $q_5 = 1.2$, $q_6 = 0.38$, in *mol/emu*
$p_2 = 14.3$, $p_3 = 48$, $p_5 = 4.4$, $p_6 = 1.3$, $q_2 = 120$, $q_3 = 26$, $q_5 = 6.9$, $q_6 = 2.1$, in $kG/\mu_B$.

Having determined the hyperfine coefficients using the NMR spin shifts and susceptibilities above $T_c$, we can now use the hyperfine coefficients and our NMR spin shifts measured also below $T_c$ to derive the susceptibilities $\chi_A$ and $\chi_B$ at all T. Instead of (5) and (6) we use the corresponding expressions for $\Delta G_\perp$ and $\Delta G_\parallel$ since this eliminates possible Meissner terms. However, we do adopt the usual definitions of our susceptibilities, $\chi_A(T=0) = 0$ and $\chi_B(T=0) = 0$. The actual susceptibility may not be zero if there is a substantial broadening of the electronic levels, which we cannot estimate. Also, although we can get $\chi_A$ and $\chi_B$ below $T_{const.}$, we cannot break the susceptibilities $\chi_A$ and $\chi_B$ up into their components $\chi_{AA}, \chi_{BB}$, and $\chi_{AB}$ below $T_{const.}$. The results are shown in figure 2 (right panel).

Recently Barzykin & Pines have published an extensive paper (Barzykin and Pines, 2009) arguing that the cuprates are two-component systems. They call one component a spin liquid (SL), the other component a Fermi liquid (FL). Their formulas are $\chi_{dd} = f(x) \cdot \chi_{SL}$ and $\chi_{pp} + 2\chi_{pd} = (1 - f(x)) \cdot \chi_{FL}$, where $x$ specifies the doping. Johnston pointed out that the temperature dependence of $F(T/T_{max})$ was the same as the theoretical form of the spin susceptibility of the spin 1/2 Heisenberg antiferromagnet. Since the antiferromagnetism arises from the Cu electron spin, and since $\chi_{AA}$ has the temperature dependence of $F(T/T_{max})$, $\chi_{AA}$ must be associated with the Cu electron spin. Therefore, our formulas support their formulation and we can identify our symbols $A$ and $B$ with their symbols $d$ and $p$, respectively. With this correspondence our results show that for optimally doped LSCO $\chi_{FL}$ is independent of temperature above $T_{const.}$ and $\chi_{SL}$ has the temperature-dependence of our $\chi_{AA}$ and thus of Johnston's $\chi_1(T)$.

Barzykin and Pines argue that there is a temperature approximately equal to $T_{max}/3$ above which $\chi_{FL}$ is independent of temperature. That temperature would be about 130K for our sample. However, our data show T-independent behavior to a much lower temperature $(T_{const.} \approx 40K)$ in the case of optimally doped $La_{1.85}Sr_{0.15}CuO_4$.

In conclusion, we have shown that a single-component description of high-temperature superconductors is not valid in general. We find that two spin components with different T-dependencies suffice to explain our data. We find that for $T \geq T_c$ in $La_{1.85}Sr_{0.15}CuO_4$ one spin component's susceptibility ($\chi_{BB}$) to be T-independent, as well as the one ($\chi_{AB}$) describing the coupling between the two components, which is negative. The pseudo-gap feature in the NMR shifts is carried by the second component's susceptibility ($\chi_{AA}$) that is T-dependent already far above $T_{const.}$

and continues to decrease through the phase transition, the point below which the first two susceptibilities ($\chi_{BB}, \chi_{AB}$) disappear rapidly. Such a two-component description seems to be able to explain various NMR shift data (Barzykin and Pines, 2009). A likely scenario is that of a planar Cu electronic spin component and another on the planar O, where the Cu spins show the pseudo-gap behavior (Johnston, 1989) and the O spins behave Pauli-like and couple to the Cu spins with a negative susceptibility.

## Acknowledgments

The authors are indebted to N. Curro, D. Pines, V. Barzykin, J. Schmalian for various discussions and suggestions with regard to the two-component approach. Financial support by the EU contract "CoMePhS", STREP No 517039, is acknowledged (JH), as well as various discussions with O.P. Sushkov (JH, GVMW).